\documentclass[11pt]{article}
\usepackage{natbib, graphicx, setspace, amsmath, amssymb, subfigure, url, multirow, booktabs, verbatim, bm,threeparttable,algorithm, color}
\usepackage{amsthm}
\usepackage{natbib}

\allowdisplaybreaks[3]
\begin{document}
	
	\setlength{\textheight}{575pt}
	\setlength{\baselineskip}{23pt}
	
	\title{Two-level Bayesian interaction analysis for survival data incorporating pathway information} 
	\author{Xing Qin$^{1}$,
		Shuangge Ma$^{2}$, and Mengyun Wu$^{1,*}$\\ \\
		$^{1}$	School of Statistics and Management, Shanghai University of Finance and Economics \\
		$^{2}$Department of Biostatistics, Yale School of Public Health\\
	\\
	email: wu.mengyun@mail.shufe.edu.cn}
	\date{} 
	\maketitle

\begin{abstract}{
	Genetic interactions play an important role in the progression of complex diseases, providing explanation of variations in disease phenotype missed by main genetic effects. Comparatively, there are fewer investigations on prognostic survival time, given its challenging characteristics such as censoring. In recent biomedical research, two-level analysis of both genes and their involved pathways has received much attention and been demonstrated to be more effective than single-level analysis, however such analysis is limited to main effects. Pathways are not isolated and their interactions have also been suggested to have important contributions to the prognosis of complex diseases. In this article, we develop a novel two-level Bayesian interaction analysis approach for survival data. This approach is the first to conduct the analysis of lower-level gene-gene interactions and higher-level pathway-pathway interactions simultaneously. Significantly advancing from existing Bayesian studies based on the Markov Chain Monte Carlo (MCMC) technique, we propose a variational inference framework based on the accelerated failure time model with favourable priors to account for two-level selection as well as censoring. The computational efficiency is much desirable for high dimensional interaction analysis. We examine performance of the proposed approach using extensive simulation. Application to TCGA melanoma and lung adenocarcinoma data leads to biologically sensible findings with satisfactory prediction accuracy and selection stability. }

{\textbf{keywords}: Bayesian analysis; Interaction analysis; Survival data; Two-level selection}
\end{abstract}

\maketitle

\section{Introduction}
\label{s:intro}

Genetic interactions have long been recognized to be fundamentally important for risk and prognosis of complex diseases \citep{power2017microbial,stange2021importance}. As a result, main-effect-only analysis has been advanced into interaction analysis to improve disease modelling \citep{2018Robust}. Representative statistical interaction analyses include \cite{Glin2015learning}, \cite{hao2018model}, and \cite{2020Sequential} based on the penalization technique, and \cite{Intro2015bayesian1}, \cite{griffin2017hierarchical}, and \cite{kim2018bayesian} based on the Bayesian technique. In these studies, much effort has been devoted to studying the ``main effects-interactions'' (M-I) hierarchy, where an interaction can be identified only when one of its corresponding main effects (weak hierarchy) or both (strong hierarchy) are identified. This hierarchy has been suggested to be useful for identification and interpretation. On the other hand, it also imposes a great challenge \citep{Intro2013a1}. Most of the aforementioned approaches are only applicable to completely observed continuous response. Interaction analysis for prognostic survival data is still limited, due to its challenging characteristics such as non-negative distributions and censoring. Existing studies include the Cox model with modified adaptive Lasso penalization \citep{WANG2014126}, where a coefficient decomposition strategy was developed for accommodating M-I hierarchy. \cite{Introwu2018identifying} developed a Cox tensor regression model with the penalization technique for identifying interactions, where the M-I hierarchy was respected using the CP decomposition.


Despite considerable successes, the above works share a common limitation that they only investigate single-level interactions between genetic factors while there are potentially complex multiple-level structures underlying genetic data. Specifically, some genes work together and are involved in the same pathway with particular cellular or physical functions. As such, there is a two-level structure, where genes are considered as the lower-level (individual) variables, and their involved pathways are considered as the higher-level (group) variables. Two-level analysis has been considered in a few recently published studies, but mostly focused on main effects only. Examples include sparse group Lasso \citep{2013sparse}, group exponential Lasso \citep{Intro2015GEL}, and adaptive sparse group Lasso \citep{2020Poignard}, which usually consider a combination of penalties at group and individual levels. In contrast, Bayesian methods often assume spike and slab priors on different levels of variables, and examples include \cite{Intro2011stingo}, \cite{Ray2016}, \cite{mallick2017bayesian}, and \cite{2018BIVAS}. There are a few other works exploring dependency structures among two-level variables, for example based on unsupervised Gaussian graphical model \citep{2019Joint}. These studies have demonstrated that two-level analysis not only improves identification and estimation accuracy but also achieves better biological interpretability. However,  methodological development of two-level analysis in the context of interaction analysis is still very limited.

\begin{figure}[H]
	\centering
	\includegraphics[width=5in]{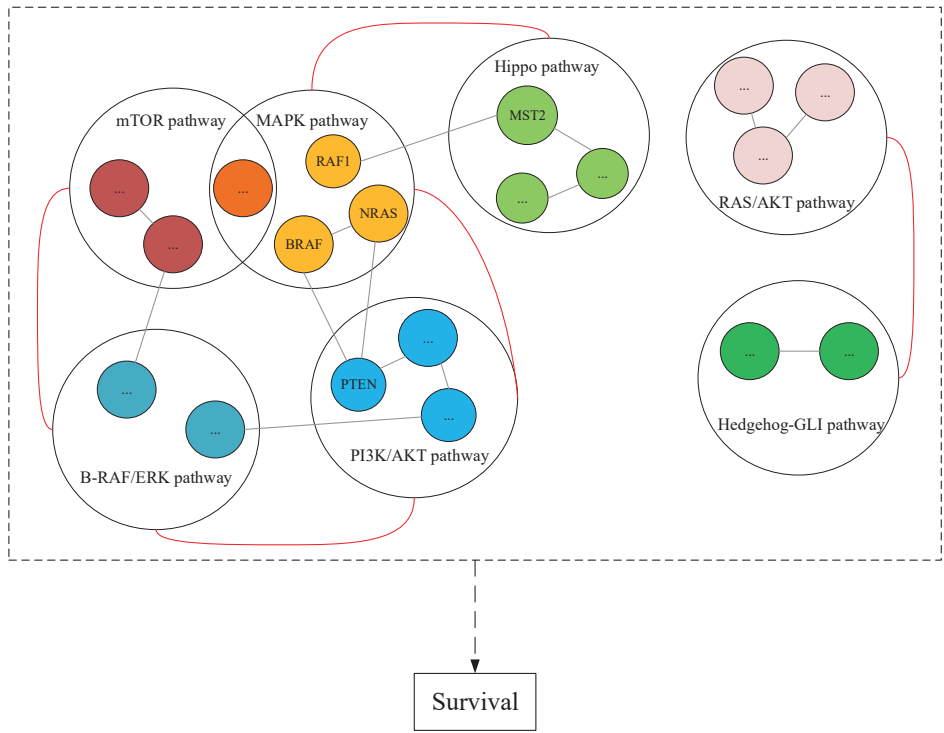}
	\caption{Partial pathway interactions for cutaneous melanoma. Different colors represent different pathways, and two genes (pathways) are connected if there is an interaction..}
	\label{fig:skcm}
\end{figure}

In recent biomedical studies, besides interactions between lower-level genes, those between higher-level pathways have also been observed, which may deliver additional information on disease development and prognosis. Take cutaneous melanoma as an example. Studies have found important pathway-pathway interactions, including the interaction of the RAS/AKT and Hedgehog-GLI pathways \citep{stecca2007melanomas}, and interactions of the B-RAF/ERK pathway with PI3K/AKT and mTOR pathways \citep{babchia2010pi3k}. In addition, \cite{feng2017mapk} has found the interaction between MAPK and Hippo pathways as well as that between their involved genes, RAF1 and MST2, in melanoma development. The PI3K/AKT and MAPK pathways have been found to interact at different stages of signal transmission, involving multiple signaling nodes and routes in cutaneous melanoma \citep{tsao2004genetic,2016Computational}. These studies have also shown that the interactions among genes NRAS, BRAF, and PTEN involved in these two pathways are critical to tumorigenesis. The graphical representation of these interactions for melanoma is provided in Figure \ref{fig:skcm}, where a two-level structure is observed. Pathway-pathway interactions have also been identified to be associated with other diseases, such as the ER and FGFR pathways interaction for non-small cell lung cancer (NSCLC) \citep{2017Interaction} and breast cancer \citep{fillmore2010estrogen}, Wnt and androgen signaling pathways interaction for prostate cancer \citep{di2017mtor}, and mTOR and PP2A pathways interaction for colorectal cancer \citep{jang2020reciprocal}. Therefore, it is of great interest to conduct two-level analysis to explore important interactions between lower-level genes and those between higher-level pathways simultaneously, for better understanding variations in disease phenotype.

In this article, we propose a new two-level Bayesian interaction analysis approach for censored survival data. Significantly advancing from existing interaction analysis \citep{hao2018model,2020Sequential}, we consider the scenarios in which interactions associated with survival have two levels: higher-level pathway-pathway interactions and lower-level gene-gene interactions. Building on the Bayesian technique, we introduce a conditional distribution based on the accelerated failure time (AFT) model with favourable priors to effectively accommodate right censoring. The proposed Bayesian AFT model enjoys simplicity, flexibility as well as
parameter interpretability not shared by Cox model-based interaction analysis with penalization \citep{WANG2014126,Introwu2018identifying}. Under this Bayesian framework, additional priors are developed to explore two-level identification for interactions as well as M-I hierarchy, which is much more challenging than two-level main-effect-only analysis \citep{2020Poignard,2018BIVAS}. We note that in existing frequentist approaches, rigid constraints or complex penalty terms are often introduced to reinforce the M-I hierarchy \citep{Intro2013a1, WANG2014126} and accommodate two-level main-effect selection \citep{Intro2015GEL,2020Poignard}. These constraints/penalties are nontrivial, and it can be conceived that further methodological development of two-level interaction analysis will be more complicated and potentially suffer from computational drawbacks. Unlike the frequentist strategy, the proposed Bayesian approach provides an intuitive hierarchical model, which is easy to interpret and produces uncertainty estimates of the coefficients, so that coherent inference can be made through the posterior distribution. In addition, different from the commonly used MCMC algorithm \citep{mallick2017bayesian,kim2018bayesian}, we develop a variational Bayesian expectation maximization algorithm to handle high dimensionality, which greatly reduces computational cost. Overall, this study provides a new systematic approach for genetic interaction analysis in prognosis modelling.

\section{Methods}
Consider $n$ independent subjects and $K$ pathways, where the $k$th pathway consists of $p_k$ genes and $\sum_{k=1}^{K}p_k=p$. For the $i$th subject, let $\boldsymbol{x}_{i}=\left(x_{i1}^{(1)},\cdots,x_{ip_{1}}^{(1)},\cdots,x_{i1}^{(k)},\right.$ $\left.\cdots,x_{ip_{k}}^{(k)},\cdots,x_{i1}^{(K)},\cdots,x_{ip_{K}}^{(K)}\right)$ be the vector of $p$ gene expression measurements, where the superscript $(k)$ indicates that the corresponding gene is involved in the $k$th pathway. Other genetic measurements, such as copy number variations, can be analyzed in the same manner. Note that if a gene is included in multiple pathways, we duplicate the corresponding measurement in these pathways. Denote $t_i$ as the survival time of subject $i$ and $y_i^{*}=\min(y_i, c_i)$ where $y_i=\log(t_i)$ and $c_i$ is the logarithm of the censoring time. We further use $\delta_i=\boldsymbol{1}_{\{y_i\leq c_i\}}$ to denote the censoring indicator, where $\boldsymbol{1}_{\{\cdot\}}$ is an indicator function. The observed data is $\left\{\left(y_i^{*}, \delta_i, \boldsymbol{x}_i\right); i=1,\cdots,n\right\}.$ Denote $\boldsymbol{X}$ as the design matrix composed of $\boldsymbol{x}_i$'s and  $\boldsymbol{\delta}$ as the vector composed of $\delta_i$'s.

\subsection{AFT Model}

For modelling survival time, we consider the Log-Normal AFT model, and the conditional density of $\log t_i$ given $\boldsymbol{x}_{i}$ has the form
\begin{equation}\label{lognormal}
	p\left(\log t_i \mid \boldsymbol{x}_{i}, \boldsymbol{w}; \tau\right)
	=\mathcal{N}\left(\log t_i \left| \sum\limits_{k=1}^{K}\sum\limits_{j=1}^{p_{k}}\left( x_{i j}^{(k)} w_{k j}^{(\boldsymbol{1})}+\sum\limits_{k^{\prime}=1}^{K}
	\sum\limits_{\substack{l>j\textrm{ if }k^{\prime}=k, \\ l=1\textrm{ if }k^{\prime}>k}}^{p_{k^{\prime}}}  x_{i j}^{(k)} x_{i l}^{\left(k^{\prime}\right)} w_{k k^{\prime}, j l}^{(\boldsymbol{2})}\right) \right., \tau^{-1}\right).
\end{equation}
where $\boldsymbol{w}\triangleq\left(w_1,\cdots,w_{(p+1)p/2}\right)^{\mathrm{T}}$ consists of all main effects $w_{kj}^{(\boldsymbol{1})}$'s and interaction effects $w_{kk',jl}^{(\boldsymbol{2})}$'s, and $\tau$ is the precision parameter.

To accommodate censoring, $\delta_i$ is assumed to follow a Binomial distribution with parameter defined from the cumulative distribution function of $y_{i}=\log t_i$. Specifically,
\begin{equation}\label{eq:delta}
	p\left(\boldsymbol{\delta}\right)=\prod_{i=1}^n\left(1-F(c_i)\right)^{1-\delta_i}F(c_i)^{\delta_i},
\end{equation}
where $F(c_i)=\int_{-\infty}^{c_i}p\left(y_i|\boldsymbol{x}_i,\boldsymbol{w};\tau\right)d y_i$. Denote $\mathcal{I}_o=\{i|\delta_i=1\}$. Then when $i\in \mathcal{I}_o$, $y_{i}^{*}\in \left(-\infty,c_i\right)$ follows a truncated Normal distribution:
\begin{equation}\label{eq:y}
	p\left(y_{i}^{*}|\delta_i=1,\boldsymbol{w};\tau\right)=\frac{1}{F(c_i)}\mathcal{N}\left(y_{i}^{*}|\widetilde{\boldsymbol{x}}_i\boldsymbol{w},\tau^{-1}\right)\boldsymbol{1}_{\{y_{i}^{*}\leq c_i\}},
\end{equation}
where $\widetilde{\boldsymbol{x}}_i$ is the vector consisting of $x_{ij}^{(k)}$'s and $x_{i j}^{(k)} x_{i l}^{\left(k^{\prime}\right)}$'s for subject $i$. When $i\in \mathcal{I}_o^c$ with $\delta_i=0$, a latent variable $z_i$ is introduced and assumed to follow a truncated distribution:
\begin{equation}\label{eq:z}
	p\left(z_i|\delta_i=0,\boldsymbol{w};\tau\right)=\frac{1}{1-F(c_i)}\mathcal{N}\left(z_i|\widetilde{\boldsymbol{x}}_i\boldsymbol{w},\tau^{-1}\right)\boldsymbol{1}_{\{z_i>c_i\}}.
\end{equation}

Thus, we obtain the joint distribution of $(\boldsymbol{y}^*,\boldsymbol{z},\boldsymbol{\delta})$ given $\boldsymbol{X}$ as
\begin{equation}\label{equation.one}
	p\left(\boldsymbol{y}^*,\boldsymbol{z},\boldsymbol{\delta}|\boldsymbol{X},\boldsymbol{w}; \tau\right)=\prod\limits_{i\in \mathcal{I}_o} p\left(y_i^{*}|\delta_i=1,\boldsymbol{x}_{i},\boldsymbol{w};\tau\right)\prod\limits_{i\in \mathcal{I}_o^{c}} p\left(z_i|\delta_i=0,\boldsymbol{x}_{i},\boldsymbol{w};\tau\right)p\left(\boldsymbol{\delta}\right),
\end{equation}
where $\boldsymbol{y}^*=\left(\{y_i^*\}_{i\in\mathcal{I}_o}\right)^{\mathrm{T}}$ and $\boldsymbol{z}=\left(\{z_i\}_{i\in\mathcal{I}_o^{c}}\right)^{\mathrm{T}}$. In (\ref{equation.one}), the main effects, and their interactions within and across pathways are jointly analyzed. The Log-Normal distribution is adopted for survival analysis, which is popular in published studies \citep{klein2003survival,2018Capturing}.
It has been observed that for many diseases including gastric carcinoma \citep{pourhoseingholi2007comparing}, breast cancer \citep{vallinayagam2014parametric}, and NSCLC \citep{claret2018model}, the Log-Normal distribution fits better than other parametric models. In addition, we can specify conjugate priors for Log-Normal distribution, drastically simplifying estimation \citep{sha2006bayesian}. We assume a Binomial distribution (\ref{eq:delta}) based on $F(c_i)$ for $\delta_i$, with the consideration that a larger $c_i$ makes it easier to obtain an uncensored observation. In addition, combining (\ref{eq:delta}) and $z_i$ introduced with prior (\ref{eq:z}), together with $y_i^*$ and its prior (\ref{eq:y}), the joint distribution of censored data becomes tractable, effectively facilitating variational inference for the posterior distribution of parameters. We note that if all subjects are observed without censoring, the AFT model (\ref{equation.one}) reduces to the ordinary Log-Normal model.

\subsection{Priors for two-level interaction analysis}

We consider that the effects (main effects and interactions) of genes associated with survival come from two levels, with the first level indicating genes' contribution and the second level indicating pathways' (higher-level gene groups) contribution. Motivated by the existing two-level analysis for main effects only \citep{Intro2011stingo,2018BIVAS}, a two-level hierarchical structure is assumed to improve interpretability, where a higher-level pathway (pathway-pathway interaction) can be selected only if at least one of its lower-level genes (gene-gene interactions) is selected, and if a lower-level gene (gene-gene interaction) is selected then its higher-level pathway (pathway-pathway interaction) should be selected. Meanwhile, we also take into account the M-I hierarchy for meaningful interaction analysis.

The following notations are first introduced:
\begin{itemize}
	\item [(1)] $\boldsymbol{\beta}=\left(\left\{ \beta_{kj}^{(\boldsymbol{1})}\right\}_{k,j} ,\left\{\beta_{kk^{\prime} ,jl}^{(\boldsymbol{2})}\right\}_{k,k',j,l} \right)^{\mathrm{T}}\triangleq\left(\beta_1,\cdots,\beta_{(p+1)p/2}\right)^{\mathrm{T}}$ is the binary lower-level selection indicator vector, with $\beta_{kj}^{(\boldsymbol{1})}$ indicating whether the $j$th main effect in the $k$th pathway is selected $\left(\beta_{kj}^{(\boldsymbol{1})}=1\right)$ or not $\left(\beta_{kj}^{(\boldsymbol{1})}=0\right)$, and $\beta_{kk^{\prime} ,jl}^{(\boldsymbol{2})}$ being that for the interaction-selection defined in a similar way.
	\item [(2)] $\boldsymbol{\alpha}=\left(\left\{\alpha_{kk'}\right\}_{k,k'}\right)^{\mathrm{T}}$ is the binary higher-level selection indicator vector, with $\alpha_{kk'}$ for identifying pathway ($k=k'$) and pathway-pathway interactions ($k\neq k'$).
	\item [(3)] $\widetilde{\boldsymbol{w}}=\left(\left\{\widetilde{\boldsymbol{w}}_{kk'}\right\}_{k,k'}\right)^{\mathrm{T}}$ is the higher-level latent vector for $\boldsymbol{w}\triangleq\left(\left\{\boldsymbol{w}_{kk'}\right\}_{k,k'}\right)^{\mathrm{T}}$, with $\boldsymbol{w}_{kk'}=\left(\left\{w_{kj}^{(\boldsymbol{1})}\right\}_{j},\left\{w_{kk,jl}^{(\boldsymbol{2})}\right\}_{j,l}\right)^{\mathrm{T}}$ if $k=k'$ and $\boldsymbol{w}_{kk'}=\left(\left\{w_{kk',jl}^{(\boldsymbol{2})}\right\}_{j,l}\right)^{\mathrm{T}}$ otherwise.
	\item [(4)] $\bar{\boldsymbol{w}}^{(\boldsymbol{2})}=\left(\left\{\bar{\boldsymbol{w}}_{kk^{\prime}}^{(\boldsymbol{2})}\right\}_{k,k'}\right)^{\mathrm{T}}$ is the lower-level latent vector for $\boldsymbol{w}^{(\boldsymbol{2})}=\left(\left\{\boldsymbol{w}_{kk^{\prime}}^{(\boldsymbol{2})}\right\}_{k,k'}\right)^{\mathrm{T}}$, with ${\boldsymbol{w}}_{kk'}^{(\boldsymbol{2})}=\left(\left\{{w}_{kk',jl}^{(\boldsymbol{2})}\right\}_{j,l}\right)^{\mathrm{T}}$ including the interaction effects within ($k=k'$) or across ($k\neq k'$) pathways.
\end{itemize}

Then, the proposed priors are defined as follows.
\begin{itemize}
	\item [(1)] Lower-level selection:
	\[p\left(w_j|\beta_j\right)=\mathcal{N}\left({w}_{ j}  | 0, r_{1}\right)^{\beta_{j} } \mathcal{N}\left({w}_{j}  | 0, r_{2}\right)^{1-\beta_{j} }\textrm{ and }   p(\beta_j)=\left(\beta_{j}\right)^{\zeta_1}\left(1-\beta_{j}\right)^{1-\zeta_1}.
	\]
	\item [(2)] Higher-level selection:
	\begin{equation*}
		p\left(\widetilde{\boldsymbol{w}}_{kk'}|\alpha_{kk'}\right)=\mathcal{N}\left(\widetilde{\boldsymbol{w}}_{kk'}  | \boldsymbol{0}, s_{1}
		{\boldsymbol{I}}\right)^{\alpha_{kk'}} \mathcal{N}\left(\widetilde{\boldsymbol{w}}_{kk'}  | \boldsymbol{0}, s_{2} \boldsymbol{I}\right)^{1-\alpha_{kk'}},
	\end{equation*}
	and
	\begin{equation*}
		p\left(\alpha_{kk'}\right)=\left(\alpha_{kk'}\right)^{\zeta_2}\left(1-\alpha_{kk'}\right)^{1-\zeta_2}.
	\end{equation*}
	\item [(3)] Two-level hierarchy:
	\[p\left(\widetilde{\boldsymbol{w}}_{kk'} |{\boldsymbol{w}}_{kk'}\right)=\boldsymbol{1}_{\{\widetilde{\boldsymbol{w}}_{kk'} ={\boldsymbol{w}}_{kk'}\}}.\]
	
	\item [(4)] M-I hierachy:
	\begin{equation*}
		\begin{aligned}
			p\left(\bar{\boldsymbol{w}}_{kk'}^{(\boldsymbol{2})}| \boldsymbol{\beta}_{k}^{(\boldsymbol{1})},\boldsymbol{\beta}_{k'}^{(\boldsymbol{1})}\right)=&\prod\limits_{j=1}^{p_{k}}\prod\limits_{\substack{l>j\textrm{ if }k^{\prime}=k, \\ l=1\textrm{ if }k^{\prime}>k}}^{p_{k^{\prime}}}\mathcal{N}\left(\bar{w}_{kk',jl}^{(\boldsymbol{2})} | 0, r_{1}\right)^{\beta_{kj}^{(\boldsymbol{1})}\beta_{k'l}^{(\boldsymbol{1})}} \mathcal{N}\left(\bar{w}_{kk',jl}^{(\boldsymbol{2})} | 0, r_{2}\right)^{1-\beta_{kj}^{(\boldsymbol{1})}\beta_{k'l}^{(\boldsymbol{1})}}
	\end{aligned}\end{equation*}
	and
	\begin{equation*}
		p\left(\bar{\boldsymbol{w}}_{kk'}^{(\boldsymbol{2})} | {\boldsymbol{w}}_{kk'} ^{(\boldsymbol{2})}\right)=\boldsymbol{1}_{\left\{\bar{\boldsymbol{w}}_{kk'} ^{(\boldsymbol{2})}={\boldsymbol{w}}_{kk'}^{(\boldsymbol{2})}\right\}}.
	\end{equation*}
\end{itemize}
Here, $r_1>r_2>0$ and $s_1>s_2>0$ are tuning parameters with $r_{2}$ and $s_2$ being close to zero, $\zeta_{1}$ and $\zeta_{2}$ are the prior lower-level and higher-level selection probabilities, and  $\boldsymbol{\beta}_k^{(\boldsymbol{1})}=\left(\beta_{k1}^{(\boldsymbol{1})},\cdots,\beta_{kp_k}^{(\boldsymbol{1})}\right)^{\mathrm{T}}$.

These priors have been motivated by the following considerations. Identification of the lower-level variables is achieved using the spike and slab prior \citep{2018BIVAS,ray2021variational}. With $r_2$ close to zero, $w_j$ is shrunk towards zero with a high probability when $\beta_j=0$. Significantly advancing from existing interaction analyses, we further introduce $\boldsymbol{w}_{kk'}$ and $\widetilde{\boldsymbol{w}}_{kk'}$ to accommodate higher-level effects, and innovatively utilize a spike and slab prior for higher-level selection with the indicator $\alpha_{kk'}$. This strategy is different from the popular two-level Bayesian variable selection studies such as \cite{zhu2019bayesian}, where each regression coefficient is reparametrized as a product of a weight parameter and an indicator, and has more intuitive interpretation and enables automatically higher-level selection. We note that reparameterization cannot guarantee the two-level hierarchy, even with additional constraints advocated by \cite{Intro2011stingo} or a conditional prior by \cite{Ray2016}, where the algorithm may become infeasible or the model may be not identifiable. In our study, we introduce a simple prior $p\left(\widetilde{\boldsymbol{w}}_{kk'} |{\boldsymbol{w}}_{kk'}\right)$ to effectively accommodate the two-level hierarchy. Specifically, when the $k$th higher-level pathway is not important ($\alpha_{kk}=0$), $\boldsymbol{w}_{kk}=\widetilde{\boldsymbol{w}}_{kk}\approx\mathbf{0}$, promoting all lower-level $\beta_{kj}^{(\boldsymbol{1})}$'s and $\beta_{kk,jl}^{(\boldsymbol{2})}$'s involved in the $k$th pathway to zero with a high probability. Moreover, if at least one of $\beta_{k j}^{(\boldsymbol{1})}$'s and $\beta_{kk,jl}^{(\boldsymbol{2})}$'s equals one, $\alpha_{kk}$ is also nonzero with a high probability. The two-level hierarchy for interactions can be discussed in a similar way. This indicator prior has another advantage that no additional intractable partition function will appear, greatly simplifying the proposed Bayesian model (we refer to Section 2.3 for more details). In a similar spirit as for the two-level hierarchy, we develop two priors for $\bar{\boldsymbol{w}}_{kk'}^{(\boldsymbol{2})}$ to accommodate the M-I hierarchy, and this strategy is more intuitive and computationally simpler than constraints/penalties-based interaction analysis. Specifically, if an interaction is selected with $\beta_{kk',jl}^{(\boldsymbol{2})}=1$, then $\bar{w}_{kk',jl}^{(\boldsymbol{2})}={w}_{kk',jl}^{(\boldsymbol{2})}\neq 0$, leading to $\beta_{kj}^{(\boldsymbol{1})}\beta_{k'l}^{(\boldsymbol{1})}=1$ (i.e. $\beta_{kj}^{(\boldsymbol{1})}=\beta_{k'l}^{(\boldsymbol{1})}=1$) with a high probability.

\subsection{Two-level Bayesian interaction analysis}

Let $\boldsymbol{\Omega}=\left\{\boldsymbol{z},\boldsymbol{w},\boldsymbol{\beta},\boldsymbol{\alpha}\right\}$ and $\boldsymbol{\Phi}=\left\{\tau,\zeta_1,\zeta_2\right\}$. The proposed Bayesian model can be formulated as
\begin{equation}\label{equation.nine}
	\begin{aligned}
		\quad&  p\left(\boldsymbol{y}^*,\boldsymbol{\delta},\boldsymbol{\Omega}|\boldsymbol{X};\boldsymbol{\Phi}\right)= p\left(\boldsymbol{y}^*,\boldsymbol{z},\boldsymbol{\delta}|\boldsymbol{X},\boldsymbol{w}; \tau\right)\left[\prod_{j=1}^{p(p+1)/2}p\left(w_j|\beta_j\right)p\left(\beta_j\right)\right] \\& \left[\prod\limits_{k=1}^{K}\prod\limits_{k'=k}^{K}p\left(\boldsymbol{w}_{kk'}|\alpha_{kk'}\right)p\left(\alpha_{kk'}\right)p\left(\boldsymbol{w}_{kk'}^{(\boldsymbol{2})}| \boldsymbol{\beta}_{k}^{(\boldsymbol{1})},\boldsymbol{\beta}_{k'}^{(\boldsymbol{1})}\right)\right].
	\end{aligned}
\end{equation}
Here we rewrite $p\left(\widetilde{\boldsymbol{w}}_{kk'}|\alpha_{kk'}\right)$ and $	p\left(\bar{\boldsymbol{w}}_{kk'}^{(\boldsymbol{2})}| \boldsymbol{\beta}_{k}^{(\boldsymbol{1})},\boldsymbol{\beta}_{k'}^{(\boldsymbol{1})}\right)$ as the generative models $p\left(\boldsymbol{0} | \widetilde{\boldsymbol{w}}_{kk'},\alpha_{kk'}\right)$ and $p\left(\boldsymbol{0} |\bar{\boldsymbol{w}}_{kk'}^{(\boldsymbol{2})},\boldsymbol{\beta}_{k}^{(\boldsymbol{1})},\boldsymbol{\beta}_{k'}^{(\boldsymbol{1})}\right)$ with observation vector $\boldsymbol{0}$. Thus in (\ref{equation.nine}), $p\left(\boldsymbol{y}^*,\boldsymbol{\delta},\boldsymbol{\Omega}|\boldsymbol{X};\boldsymbol{\Phi}\right)$ can be reformulated as a hybrid Bayesian model with tractable partition functions. In addition, with $p\left(\widetilde{\boldsymbol{w}}_{kk'} |{\boldsymbol{w}}_{kk'}\right)$ and $p\left(\bar{\boldsymbol{w}}_{kk'}^{(\boldsymbol{2})} | {\boldsymbol{w}}_{kk'} ^{(\boldsymbol{2})}\right)$ being indicator functions,  $\widetilde{\boldsymbol{w}}_{kk'}$ and $\bar{\boldsymbol{w}}_{kk'}^{(\boldsymbol{2})}$ can be replaced by their counterparts of $\boldsymbol{w}$. Details of (\ref{equation.nine}) are provided in Appendix A of the Supplementary Materials.

Denote $\textrm{E}(\beta_{j})$ and $\textrm{E}(\alpha_{kk'})$ as the posterior expectations of the lower-level and higher-level selection indicators $\beta_{j}$ and $\alpha_{kk'}$ under (\ref{equation.nine}).
Like \citet{ray2021variational}, we identify lower-level and higher-level variables with $\textrm{E}(\beta_{j})$'s and $\textrm{E}(\alpha_{kk'})$'s greater than 0.5 as important.

\subsection{Computation}

To obtain the posterior distribution $p\left( \boldsymbol{\Omega} \mid \boldsymbol{y}^{*}, \boldsymbol{\delta},\boldsymbol{X} ; \boldsymbol{\Phi}\right)$ and estimates of $\boldsymbol{\Phi}$, we develop an effective variational Bayesian expectation maximization (VBEM) algorithm, where a computationally tractable distribution is estimated to approximate the intractable $p\left( \boldsymbol{\Omega} \mid \boldsymbol{y}^{*}, \boldsymbol{\delta},\boldsymbol{X} ; \boldsymbol{\Phi}\right)$. Denote  $q\left(\boldsymbol{\Omega}\right)$ as a candidate approximating distribution of $p\left( \boldsymbol{\Omega} \mid \boldsymbol{y}^{*}, \boldsymbol{\delta},\boldsymbol{X} ; \boldsymbol{\Phi}\right)$ and $E_q$ as the expectation taken with respect to $q\left(\boldsymbol{\Omega}\right)$. We decompose the log-marginal likelihood as
$$
\log p(\boldsymbol{y}^{*},\boldsymbol{\delta} \mid \boldsymbol{X} ; \boldsymbol{\Phi})=\mathcal{L}_{q}+\operatorname{KL}\left(q(\boldsymbol{\Omega}) \|p\left( \boldsymbol{\Omega} \mid \boldsymbol{y}^{*}, \boldsymbol{\delta},\boldsymbol{X} ; \boldsymbol{\Phi}\right)\right),
$$
where
$$
\begin{aligned}
	\mathcal{L}_{q} &=\mathrm{E}_{q} \log \left[\frac{p(\boldsymbol{y}^{*},\boldsymbol{\delta}, \boldsymbol{\Omega}\mid \boldsymbol{X} ; \boldsymbol{\Phi})}{q(\boldsymbol{\Omega})}\right], \\
	\mathrm{KL}\left(q(\boldsymbol{\Omega})\|p(\boldsymbol{\Omega} | \boldsymbol{y}^{*},\boldsymbol{\delta}, \boldsymbol{X};\boldsymbol{\Phi})\right)&=\int q(\boldsymbol{\Omega})\log \left[\frac{ q(\boldsymbol{\Omega})}{p(\boldsymbol{\Omega}| \boldsymbol{y}^{*},\boldsymbol{\delta}, \boldsymbol{X}; \boldsymbol{\Phi})}\right]\, d\boldsymbol{\Omega}.
\end{aligned}
$$

In the expectation (E) step, the proposed algorithm minimizes the KL divergence with respect to the approximating distribution $q(\boldsymbol{\Omega})$, which is
equivalent to maximize the evidence lower bound $\mathcal{L}_q$. To improve computational feasibility, we propose using a mean field variational family of $q(\boldsymbol{\Omega})$, where $q(\boldsymbol{\Omega})$ has the tractable form
$q\left(\boldsymbol{\Omega}\right)=q\left(\boldsymbol{w}\right)q\left(\boldsymbol{\beta}\right) q\left(\boldsymbol{\alpha}\right)q\left(\boldsymbol{z}\right)$. Following \cite{2018BIVAS}, we further factorize $q\left(\boldsymbol{w}\right)$ and $q\left(\boldsymbol{\beta}\right)$ as the products of variational distributions of $w_j$'s and $\beta_j$'s $(j=1,2,\cdots,p(p+1)/2)$, respectively, and obtain the optimal solution of $q(\boldsymbol{\Omega})$ in a similar manner with the coordinate decent algorithm. As demonstrated in \cite{blei2017variational}, although the mean-field family assumption cannot capture the correlation between latent variables, it can facilitate calculations to a certain extent and has satisfactory numerical results.  Specifically, the optimal variational distributions are estimated as
\begin{equation*}
	\begin{aligned}
		q(\boldsymbol{w})=&\prod_{j=1}^{(p+1)p/2}\mathcal{N}\left(w_{j}\left| m_{j}, \sigma_{j}^{2}\right.\right) , \quad q(\boldsymbol{\beta})=\prod_{j=1}^{(p+1)p/2} (\eta_{j})^{\beta_{j}}\left(1-\eta_{j}\right)^{1-\beta_{j}},
		\\
		q(\boldsymbol{\alpha})=&\prod_{k=1}^{K}\prod_{k'=k}^K \left(r_{kk'}\right)^{\alpha_{kk'}}\left(1-r_{kk'}\right)^{1-\alpha_{kk'}},\quad q\left(\boldsymbol{z}\right)=\prod\limits_{i\in \mathcal{I}_o^{c}}\frac{\boldsymbol{1}_{\{z_i>c_i\}}}{1-F(c_i)}\mathcal{N}\left(z_i|\widetilde{\boldsymbol{x}}_i\boldsymbol{m},\tau^{-1}\right),
	\end{aligned}
\end{equation*}
where
$\left(m_j, \sigma_j\right)$ are the corresponding estimated values of the parameters for the Normal distributions,  $\left(\eta_{j},r_{kk'}\right)$ are the expectations of $\left(\beta_{j},\alpha_{kk'}\right)$ under $q(\boldsymbol{\Omega})$, and $\boldsymbol{m}=(m_1,\cdots,m_{p(p+1)/2})^T$. The detailed derivations and estimations of $m_j$'s, $\sigma_j$'s, $\eta_j$'s, and $r_{kk'}$'s are provided in Appendix A of the Supplementary Material.

In the maximization (M) step, the proposed algorithm obtains $\mathcal{L}_q$ by calculating the expectation with respect to the approximate distribution  $q(\boldsymbol{\Omega})$. Then the current $\mathcal{L}_q$ is optimized with respect to the model parameters $\boldsymbol{\Phi}$. With some derivations, we have
\begin{equation*}\label{equation.ten}
	\begin{aligned}
		\tau=    & \frac{n_1}{\sum_{i\in \mathcal{I}_o}\left(\left(y_i^{*}\right)^2-2\widetilde{\boldsymbol{x}}_i \boldsymbol{m}y_i^{*}+\left(\widetilde{\boldsymbol{x}}_i\boldsymbol{m}\right)^2  \right)+\sum_{i}^n\widetilde{\boldsymbol{x}}_i\operatorname{diag}\left(\boldsymbol{\sigma}^2\right)\widetilde{\boldsymbol{x}}_i^{\mathrm{T}}},
		\\ \zeta_1= & \frac{\sum_{j=1}^{(p+1)p/2}\eta_j}{p(p+1)/2},\quad
		\zeta_2= \frac{ \sum_{k=1}^{K}\sum_{k'=k}^{K}r_{kk'}}{K(K+1)/2},
	\end{aligned}
\end{equation*}
where $n_1=\sum_{i=1}^n\boldsymbol{1}_{\{\delta_i=1\}}$ and $\boldsymbol{\sigma}^2=(\sigma_1^2,\cdots,\sigma_{p(p+1)/2}^2)^T$. The E and M steps are conducted iteratively until convergence, and the final estimated values of $\eta_{j}$ and $r_{kk'}$ are adopted as the estimators of $\textrm{E}\left(\beta_{j}\right)$ and $\textrm{E}\left(\alpha_{kk'}\right)$. We refer to Appendix A in the Supplementary Material for details of the VBEM algorithm. As a variational extension of the EM algorithm, the VBEM algorithm guarantees that the lower bound of the marginal likelihood increases (or remains unchanged) at each step, thus convergence of the algorithm is promised \citep{blei2017variational,2018BIVAS}.

The proposed VBEM algorithm advances from the commonly adopted MCMC algorithm in multiple aspects. First, the MCMC algorithm often falls into a local support of the target distribution without proper initial samples or transition probability density. In addition, with high dimensional parameters, it usually needs many times of cyclic sampling for MCMC to converge, leading to tremendous computational cost. However, the proposed VBEM algorithm transforms complex inference problems into high-dimensional optimization problems, and the family of approximating distribution $q(\cdot)$ controls the complexity of this optimization. As long as $q(\cdot)$ is flexible and simple enough for efficient optimization, efficient computational properties and satisfactory estimation performance can be achieved.

There are four tuning parameters: $s_{1}$, $s_{2}$, $r_{1}$ and $r_{2}$. In our numerical studies, we find that the proposed approach is not sensitive to the choice of $s_1$ and $r_1$ when they are in a sensible range, and we fix $s_1=r_1=1$.
The high speed of the VBEM algorithm allows us to consider a set of models based on different values of $s_2$ and $r_2$, and we use the Bayesian information criterion (BIC) to pick the best one. Take a simulated dataset with $p=1,000$ and $n=400$ as an example. With fixed tuning parameters, it takes about half a minute to perform the proposed analysis using a laptop with standard configurations. The proposed approach has been implemented using an R package \textit{survInter} which is available at https://github.com/mengyunwu2020/survInter.

\section{Simulation}
We conduct simulation to investigate performance of the proposed approach under the following settings. (a) $n=400$. (b) There are 1,000 distinct genes, among which 22 are involved in 2 to 6 pathways, and others are involved in only one pathway. (c) Two settings for the number of pathways are considered with $K=100$ and $50$, where the sizes of pathways are from 10 to 13 and from 20 to 23, respectively.
(d) $\boldsymbol{x}_i, i=1,\cdots,n$ are simulated independently from a multivariate Normal distribution with marginal zero means and unit variances. Two correlation structures are considered. The first one assumes an AR (autoregressive) correlation structure within each pathway, where the correlation coefficient between the $j$th and $j'$th gene is $\rho^{|j-j'|}$, and a zero correlation across pathways. Thus, distinct genes in different pathways are independent from each other. Two settings AR(0.6) and AR(0.4) with $\rho=0.6$ and $0.4$ are examined. The second is a confounding group setting (CR) introduced in \cite{yang2020consistent}. Specifically, the correlation between a gene in the active pathways and that in one of the inactive pathways (confounding pathways) is set to be 0.1, and the other across-pathway correlations are equal to 0. Two settings for within-pathway correlation are considered. Under the first setting (CR1), correlations between genes within each pathway are 0.2, and under the second setting (CR2), we consider an AR correlation structure with $\rho=0.6$. (e) Four pathways and two pathway-pathway interactions are important, containing 20 nonzero lower-level main effects and 24 nonzero lower-level gene-gene interactions. Both the two-level and M-I hierarchies are satisfied. All nonzero coefficients are generated from Uniform(0.8,1.2). Three specific settings S1-S3 for important variables are considered. Under setting S1, all important genes have a positive impact on survival time. Setting S2 is the same as S1, except that the signals for one pathway and those for one pathway-pathway interaction are negative. Under setting S3, within each pathway, the signals can be either positive or negative. We refer to Appendix B in the Supplementary Material for more details. (f) For the survival response, we first generate $y_i$ from the Normal distribution (\ref{lognormal}) with unit variance, and the survival time $t_i=e^{y_i}$. The censoring time is simulated from a Gamma distribution. By changing the shape and scale parameters, the censoring rate is controlled to be either 20$\%$ or 40$\%$. There are 42 scenarios, comprehensively covering a wide spectrum with different patterns of pathways, and different levels of correlations within and across pathways, signals associated with the response, and censoring rates (we note that under setting CR2 with $K=50$, the covariance matrix is singular, and the corresponding scenarios are not available.)

Beyond the proposed approach, we also conduct five alternatives. (a) surBayes, a survival Bayesian approach based on model (\ref{equation.one}) with only lower-level spike and slab prior on the main effects and interactions. (b) glinternet, a method for learning pairwise interactions with strong hierarchy via the group-Lasso, which is implemented using  R package \textit{glinternet} \citep{Glin2015learning}. (c) RAMP, an efficient interaction analysis algorithm
under marginality principle for computing a hierarchy-preserving regularization solution path, which can be realized using R package \textit{RAMP} \citep{hao2018model}. (d) Cox-GEL, a two-level penalization approach based on Cox model and group exponential Lasso \citep{Intro2015GEL}, realized using R package \textit{grpreg}.
(e) Cox-Lasso, the Cox model with Lasso penalty, which can be realized using R package \textit{glmnet}. Among these alternatives, surBayes is the most direct competitor but does not accommodate higher-level selection and M-I hierarchy. Glinternet and RAMP are also able to respect the strong M-I hierarchy but only explore single-level interactions. These two approaches have been developed for completely observed continuous response and cannot be directly applied to censored survival time. Following \cite{Hao2019INFERENCE}, we consider the extensions of glinternet and RAMP based on the AFT model and weighted method in our numerical studies (details are provided in Appendix B of the Supplementary Material). Cox-GEL and Cox-Lasso are applied to the stacked main effects and all pairwise interactions directly without accounting for the M-I hierarchy. In addition, Cox-GEL can also conduct two-level selection. We choose these alternatives as they have publicly available packages and satisfactory performance.

For evaluating lower-level identification performance, we adopt the numbers of true positives and false positives for main effects (L-M:TP and L-M:FP) and interactions (L-I:TP and L-I:FP). For the proposed approach and Cox-GEL, we also compute the true positives (H-M:TP and H-I:TP) and false positives (H-M:FP and H-I:FP) for identifying higher-level pathways and pathway-pathway interactions. Denote
$(\hat{\boldsymbol{w}}_{\mathcal{M}},\hat{\boldsymbol{w}}_{\mathcal{I}})$ and $(\boldsymbol{w}_{\mathcal{M}}^{0},\boldsymbol{w}_{\mathcal{I}}^{0})$ as the estimated and true values of coefficients for main effects and interactions. To assess estimation performance, we consider their root sum of squared errors M:RSSE and I:RSSE, which are defined as $||\hat{\boldsymbol{w}}_{\mathcal{M}}-\boldsymbol{w}_{\mathcal{M}}^{0}||_{2}$ and $||\hat{\boldsymbol{w}}_{\mathcal{I}}-\boldsymbol{w}_{\mathcal{I}}^{0}||_{2}$, respectively. In addition, prediction accuracy is examined based on 100 independent testing subjects using the C-statistic, which is realized using R function \textit{UnoC} and with a larger value indicating better prediction \citep{uno2011c}.

For each scenario, we repeat simulation 100 times. The lower-level results under the scenarios with correlation structure AR(0.6) are summarized in Table \ref{t:sim1} ($K=100)$ and Table \ref{t:sim2} ($K=50$). The corresponding higher-level identification results for the proposed approach and Cox-GEL are provided in Table \ref{sim:t3}. The rest of the results are provided in Appendix B of the Supplementary Material. Under all scenarios, the proposed approach is observed to perform better than the alternatives in terms of lower-level identification. For instance, under the scenario with setting S1 and censoring rate $20\%$ in Table \ref{t:sim1}, the proposed approach has (L-M:TP, L-M:FP, L-I:TP, L-I:FP) = (18.00, 0.04, 21.36, 0.16), compared to (18.18, 37.76, 12.04, 24.88) for surBayes, (18.80,  14.96, 16.94, 14.74) for glinternet, (17.04, 10.08, 14.94, 197.38) for RAMP, (15.22, 21.80, 13.86, 244.46) for Cox-GEL, and (11.14, 0.08, 13.34, 20.60) for Cox-Lasso. Despite a slight advantage of surBayes and glinternet in L-M:TP, they are far inferior in L-I:TP, L-M:FP, and L-I:FP. Under the scenarios with more complex associations with the survival time and a higher censoring rate, the proposed approach has even better performance. For example, under the scenario with S3 and $40\%$ censoring in Table \ref{t:sim1}, we observe that (L-M:TP, L-M:FP, L-I:TP, L-I:FP) of the proposed approach is (17.34, 0.16, 16.14, 2.26), compared to
(10.78, 86.70, 4.54, 32.62) for surBayes, (9.04, 20.30, 3.96, 14.36) for glinternet, (11.28,11.34, 6.34, 112.82) for RAMP, (7.48, 11.60, 6.00, 106.82) for Cox-GEL, and (5.68, 0.18, 7.22, 32.34) for Cox-Lasso. Satisfactory higher-level identification performance of the proposed approach is also observed. Under all scenarios, the proposed approach can identify the majority of important higher-level variables with almost zero false positives. Take the scenario in Table \ref{sim:t3} with S1, $K=100$ and $40\%$ censoring as an example, the values of (H-M:TP, H-M:FP, H-I:TP, H-I:FP) are (3.92, 0.00, 1.98, 0.00) for the proposed approach and (2.00, 0.00, 0.12, 0.06) for Cox-GEL. In addition, the proposed approach has higher estimation and prediction accuracy. For example, under setting S2 with $20\%$ censoring in Table \ref{t:sim2}, the proposed approach has (M:RSSE, I:RSSE, C-statistic)=(1.20, 1.50, 0.93), compared to (2.97, 4.35, 0.70) for surBays, (2.20, 38.29, 0.74) for glinternet, (4.97, 14.48, 0.64) for RAMP, (33.44, 92.26, 0.60) for Cox-GEL, and (4.12, 4.95, 0.79) for Cox-Lasso. Under the scenarios with other correlation structures, including the two confounding group structures (Tables S1-S8 in Appendix B of the Supplementary Materials), similar patterns are observed.

\begin{table}[H]
	\caption{Simulation results under the scenarios with AR(0.6) and $K=100$. In each cell, mean (SD) based on 100 replicates.}
	\renewcommand\tabcolsep{1.6pt}
	\begin{tabular}{@{}llllllll@{}}
		\toprule
		\textbf{Approach} & \textbf{L-M:TP} & \textbf{L-M:FP} & \textbf{M:RSSE} & \textbf{L-I:TP} & \textbf{L-I:FP} & \textbf{I:RSSE} & \textbf{C-statistic} \\ \midrule
		\multicolumn{8}{c}{S1 with censoring rate 40$\%$}                                                            \\
		proposed   & 15.26(2.35) & 0.10(0.30)   & 2.20(0.48)   & 17.54(3.03) & 0.68(1.06)     & 2.62(0.63)   & 0.88(0.04)  \\
		surBayes   & 15.50(2.43) & 28.02(6.14)  & 3.31(0.14)   & 8.96(2.71)  & 20.34(13.71)   & 4.48(0.17)   & 0.68(0.05)  \\
		glinternet & 17.66(1.71) & 12.70(8.19)  & 2.45(0.31)   & 13.40(4.31) & 13.38(5.77)    & 32.06(6.42)  & 0.71(0.05)  \\
		RAMP       & 15.10(1.97) & 9.08(2.47)   & 4.90(1.36)   & 10.74(5.51) & 135.36(32.73)  & 12.54(4.07)  & 0.68(0.06)  \\
		Cox-GEL    & 9.70(5.57)  & 14.06(7.36)  & 12.08(15.97) & 8.26(4.60)  & 144.02(78.22)  & 28.93(44.98) & 0.61(0.07)  \\
		Cox-Lasso  & 7.62(1.76)  & 0.12(0.33)   & 4.08(0.04)   & 9.00(2.25)  & 29.04(4.02)    & 4.89(0.04)   & 0.74(0.05)  \\ \midrule
		\multicolumn{8}{c}{S2 with censoring rate 40$\%$}                                                            \\
		proposed   & 15.10(3.25) & 0.12(0.39)   & 2.19(0.62)   & 17.26(3.78) & 0.60(1.01)     & 2.65(0.63)   & 0.87(0.13)  \\
		surBayes   & 15.90(2.51) & 27.12(6.38)  & 3.26(0.18)   & 8.82(2.71)  & 21.08(14.16)   & 4.47(0.15)   & 0.70(0.05)  \\
		glinternet & 17.62(2.69) & 12.92(7.14)  & 2.37(0.42)   & 14.26(5.14) & 13.92(6.40)    & 32.72(9.52)  & 0.72(0.04)  \\
		RAMP       & 15.42(2.23) & 8.88(2.16)   & 4.81(1.27)   & 12.22(5.79) & 138.52(30.15)  & 13.20(3.93)  & 0.69(0.08)  \\
		Cox-GEL    & 10.28(5.63) & 14.50(7.30)  & 14.35(17.70) & 9.02(5.10)  & 157.54(97.93)  & 32.94(45.86) & 0.62(0.08)  \\
		Cox-Lasso  & 7.82(1.88)  & 0.12(0.33)   & 4.08(0.04)   & 9.00(2.56)  & 27.78(5.10)    & 4.89(0.04)   & 0.75(0.05)  \\ \midrule
		\multicolumn{8}{c}{S3 with censoring rate 40$\%$}                                                            \\
		proposed   & 17.34(1.78) & 0.16(0.37)   & 1.95(0.47)   & 16.14(2.64) & 2.26(2.35)     & 2.97(0.43)   & 0.88(0.06)  \\
		surBayes   & 10.78(3.43) & 86.70(53.18) & 3.80(0.20)   & 4.54(1.68)  & 32.62(27.87)   & 4.72(0.10)   & 0.59(0.12)  \\
		glinternet & 9.04(5.31)  & 20.30(14.18) & 3.55(0.42)   & 3.96(3.50)  & 14.36(8.28)    & 14.06(5.95)  & 0.62(0.12)  \\
		RAMP       & 11.28(3.77) & 11.34(2.88)  & 4.60(1.16)   & 6.34(5.11)  & 112.82(50.58)  & 8.96(2.73)   & 0.62(0.12)  \\
		Cox-GEL    & 7.48(3.38)  & 11.60(5.12)  & 6.89(8.58)   & 6.00(2.71)  & 106.82(47.10)  & 12.88(21.96) & 0.62(0.12)  \\
		Cox-Lasso  & 5.68(1.90)  & 0.18(0.48)   & 4.06(0.05)   & 7.22(1.68)  & 32.34(4.99)    & 4.92(0.04)   & 0.74(0.11)  \\ \midrule
		\multicolumn{8}{c}{S1 with censoring rate 20$\%$}                                                            \\
		proposed   & 18.00(1.54) & 0.04(0.28)   & 1.36(0.58)   & 21.36(1.91) & 0.16(0.55)     & 1.65(0.62)   & 0.93(0.02)  \\
		surBayes   & 18.18(1.64) & 37.76(6.82)  & 2.87(0.18)   & 12.04(1.82) & 24.88(12.07)   & 4.12(0.20)   & 0.74(0.03)  \\
		glinternet & 18.80(1.43) & 14.96(7.22)  & 2.14(0.32)   & 16.94(3.94) & 14.74(5.52)    & 39.58(5.72)  & 0.73(0.04)  \\
		RAMP       & 17.04(1.67) & 10.08(2.32)  & 4.31(1.53)   & 14.94(5.53) & 197.38(42.14)  & 12.87(5.15)  & 0.71(0.06)  \\
		Cox-GEL    & 15.22(5.77) & 21.80(7.84)  & 23.46(21.64) & 13.86(6.12) & 244.46(120.96) & 54.75(54.04) & 0.68(0.08)  \\
		Cox-Lasso  & 11.14(1.90) & 0.08(0.27)   & 4.13(0.04)   & 13.34(1.98) & 20.60(3.98)    & 4.95(0.04)   & 0.80(0.04)  \\ \midrule
		\multicolumn{8}{c}{S2 with censoring rate 20$\%$}                                                            \\
		proposed   & 17.76(1.76) & 0.02(0.14)   & 1.43(0.59)   & 21.06(1.98) & 0.20(0.57)     & 1.76(0.63)   & 0.92(0.03)  \\
		surBayes   & 18.56(1.43) & 38.08(8.96)  & 2.87(0.19)   & 12.30(1.85) & 27.10(13.35)   & 4.10(0.22)   & 0.74(0.04)  \\
		glinternet & 19.10(1.18) & 17.30(10.11) & 2.03(0.37)   & 17.88(3.94) & 16.38(6.95)    & 40.14(6.98)  & 0.74(0.03)  \\
		RAMP       & 17.08(1.54) & 9.40(2.39)   & 4.07(1.40)   & 13.64(5.69) & 186.86(46.81)  & 13.09(4.71)  & 0.70(0.07)  \\
		Cox-GEL    & 15.70(5.87) & 22.68(8.36)  & 24.76(20.94) & 14.20(6.05) & 248.78(119.20) & 57.90(50.67) & 0.68(0.07)  \\
		Cox-Lasso  & 11.40(2.13) & 0.10(0.30)   & 4.13(0.04)   & 13.72(2.19) & 19.76(4.73)    & 4.95(0.04)   & 0.80(0.03)  \\ \midrule
		\multicolumn{8}{c}{S3 with censoring rate 20$\%$}                                                            \\
		proposed   & 18.52(1.13) & 0.02(0.14)   & 1.36(0.42)   & 18.78(2.28) & 1.26(1.45)     & 2.49(0.41)   & 0.90(0.03)  \\
		surBayes   & 14.46(1.64) & 25.68(7.47)  & 3.31(0.15)   & 5.72(2.03)  & 9.24(4.24)     & 4.53(0.09)   & 0.67(0.05)  \\
		glinternet & 13.82(2.80) & 14.30(5.61)  & 3.15(0.27)   & 7.96(2.96)  & 11.32(3.46)    & 23.10(6.24)  & 0.69(0.06)  \\
		RAMP       & 16.04(1.98) & 10.64(2.67)  & 3.21(1.10)   & 13.22(6.03) & 185.28(63.10)  & 8.70(2.97)   & 0.71(0.08)  \\
		Cox-GEL    & 11.28(5.77) & 17.14(8.36)  & 10.93(14.17) & 9.38(5.11)  & 166.58(92.88)  & 21.48(33.23) & 0.66(0.07)  \\
		Cox-Lasso  & 8.84(1.75)  & 0.08(0.27)   & 4.12(0.04)   & 10.00(1.90) & 26.44(3.99)    & 4.95(0.04)   & 0.77(0.07)  \\ \bottomrule
	\end{tabular}\label{t:sim1}
\end{table}

\begin{table}[H]
	\caption{Simulation results under the scenarios with AR(0.6) and $K=50$. In each cell, mean (SD) based on 100 replicates.}
	\renewcommand\tabcolsep{1.4pt}
	\begin{tabular}{@{}llllllll@{}}
		\toprule
		\textbf{Approach} & \textbf{L-M:TP} & \textbf{L-M:FP} & \textbf{M:RSSE} & \textbf{L-I:TP} & \textbf{L-I:FP} & \textbf{I:RSSE} & \textbf{C-statistic}\\ \midrule
		\multicolumn{8}{c}{S1 with censoring rate 40$\%$}                                                           \\
		proposed   & 16.22(3.32) & 0.66(0.92)   & 1.94(0.70)   & 17.58(4.86) & 7.62(5.92)     & 2.57(0.91)    & 0.87(0.07)  \\
		surBayes   & 15.66(2.70) & 28.00(7.27)  & 3.31(0.23)   & 8.06(2.47)  & 19.32(13.53)   & 4.54(0.11)    & 0.66(0.07)  \\
		glinternet & 17.42(3.32) & 12.78(7.55)  & 2.42(0.52)   & 13.62(4.97) & 13.74(6.95)    & 31.23(8.30)   & 0.71(0.08)  \\
		RAMP       & 11.16(2.19) & 12.28(2.60)  & 5.37(1.09)   & 9.82(4.90)  & 116.58(39.91)  & 11.76(4.11)   & 0.65(0.06)  \\
		Cox-GEL    & 6.24(3.51)  & 20.54(11.28) & 23.28(15.75) & 5.38(2.65)  & 321.88(164.04) & 85.72(45.78)  & 0.56(0.08)  \\
		Cox-Lasso  & 7.36(1.95)  & 0.08(0.27)   & 4.07(0.05)   & 8.32(2.39)  & 29.56(5.07)    & 4.88(0.05)    & 0.71(0.10)  \\ \midrule
		\multicolumn{8}{c}{S2 with censoring rate 40$\%$}                                                           \\
		proposed   & 15.36(3.23) & 0.96(1.44)   & 2.17(0.58)   & 16.14(4.18) & 8.38(4.83)     & 2.94(0.71)    & 0.85(0.11)  \\
		surBayes   & 15.44(2.48) & 27.56(6.41)  & 3.31(0.18)   & 7.96(2.59)  & 18.64(14.01)   & 4.52(0.11)    & 0.67(0.05)  \\
		glinternet & 17.00(3.76) & 13.86(7.91)  & 2.46(0.52)   & 12.90(5.11) & 13.86(6.79)    & 31.08(9.37)   & 0.70(0.14)  \\
		RAMP       & 11.32(1.85) & 11.86(2.34)  & 5.44(1.21)   & 9.54(4.33)  & 122.94(41.20)  & 13.63(4.80)   & 0.64(0.08)  \\
		Cox-GEL    & 7.50(4.05)  & 23.98(12.35) & 27.47(16.52) & 6.28(2.96)  & 355.70(161.80) & 88.89(43.06)  & 0.56(0.07)  \\
		Cox-Lasso  & 7.66(2.19)  & 0.14(0.40)   & 4.07(0.05)   & 8.48(2.37)  & 29.78(5.16)    & 4.88(0.04)    & 0.72(0.06)  \\ \midrule
		\multicolumn{8}{c}{S3 with censoring rate 40$\%$}                                                           \\
		proposed   & 16.50(2.60) & 1.56(1.74)   & 2.17(0.58)   & 13.04(3.56) & 8.42(6.13)     & 3.49(0.57)    & 0.81(0.12)  \\
		surBayes   & 11.50(2.88) & 91.46(53.64) & 3.79(0.18)   & 4.24(1.41)  & 32.38(22.50)   & 4.74(0.09)    & 0.57(0.11)  \\
		glinternet & 8.84(5.56)  & 20.02(6.68)  & 3.57(0.41)   & 3.42(3.33)  & 13.04(2.42)    & 13.18(6.14)   & 0.61(0.14)  \\
		RAMP       & 8.70(2.85)  & 12.70(3.48)  & 5.45(1.31)   & 5.42(4.80)  & 106.64(49.02)  & 11.30(5.18)   & 0.56(0.14)  \\
		Cox-GEL    & 6.20(2.67)  & 21.04(8.42)  & 30.06(14.57) & 4.94(2.11)  & 287.10(115.89) & 104.08(46.68) & 0.55(0.12)  \\
		Cox-Lasso  & 5.96(1.94)  & 0.08(0.27)   & 4.06(0.04)   & 6.62(1.66)  & 33.08(5.07)    & 4.91(0.04)    & 0.70(0.12)  \\ \midrule
		\multicolumn{8}{c}{S1 with censoring rate 20$\%$}                                                           \\
		proposed   & 18.70(1.16) & 0.38(0.99)   & 1.16(0.56)   & 21.66(2.10) & 2.62(3.33)     & 1.54(0.73)    & 0.93(0.03)  \\
		surBayes   & 18.26(1.72) & 37.58(8.38)  & 2.95(0.21)   & 10.18(1.73) & 25.74(14.95)   & 4.37(0.13)    & 0.70(0.04)  \\
		glinternet & 18.98(2.28) & 16.80(7.87)  & 2.06(0.46)   & 17.18(4.54) & 16.36(5.89)    & 39.96(6.62)   & 0.74(0.05)  \\
		RAMP       & 12.28(1.51) & 13.66(2.43)  & 5.16(1.40)   & 10.90(4.06) & 177.12(59.51)  & 13.71(5.93)   & 0.65(0.05)  \\
		Cox-GEL    & 11.88(5.89) & 37.04(19.83) & 34.77(20.89) & 9.76(5.14)  & 540.70(328.74) & 89.05(55.97)  & 0.62(0.07)  \\
		Cox-Lasso  & 11.08(2.04) & 0.16(0.37)   & 4.13(0.05)   & 12.36(2.11) & 21.24(4.37)    & 4.94(0.05)    & 0.79(0.03)  \\ \midrule
		\multicolumn{8}{c}{S2 with censoring rate 20$\%$}                                                           \\
		proposed   & 18.60(1.28) & 0.40(0.78)   & 1.20(0.49)   & 21.94(1.85) & 2.82(3.01)     & 1.50(0.62)    & 0.93(0.03)  \\
		surBayes   & 18.22(1.54) & 38.52(8.24)  & 2.97(0.16)   & 10.36(1.56) & 25.06(10.92)   & 4.35(0.10)    & 0.70(0.04)  \\
		glinternet & 18.36(1.79) & 16.02(6.57)  & 2.20(0.46)   & 16.08(4.41) & 15.14(5.28)    & 38.29(8.05)   & 0.74(0.04)  \\
		RAMP       & 12.24(1.76) & 13.24(2.86)  & 4.97(1.26)   & 9.56(5.20)  & 170.14(58.15)  & 14.48(6.19)   & 0.64(0.06)  \\
		Cox-GEL    & 10.58(4.99) & 33.12(16.15) & 33.44(20.23) & 8.96(4.42)  & 493.96(280.31) & 92.26(57.39)  & 0.60(0.06)  \\
		Cox-Lasso  & 10.92(2.17) & 0.14(0.35)   & 4.12(0.04)   & 12.72(2.20) & 21.22(5.16)    & 4.95(0.05)    & 0.79(0.04)  \\ \midrule
		\multicolumn{8}{c}{S3 with censoring rate 20$\%$}                                                           \\
		proposed   & 18.66(1.06) & 1.58(2.02)   & 1.43(0.43)   & 17.38(2.60) & 6.52(6.08)     & 2.72(0.51)    & 0.87(0.04)  \\
		surBayes   & 14.56(1.50) & 27.26(7.66)  & 3.31(0.13)   & 5.52(1.72)  & 7.92(4.71)     & 4.54(0.10)    & 0.67(0.07)  \\
		glinternet & 13.92(2.39) & 14.58(4.07)  & 3.23(0.30)   & 7.02(2.65)  & 11.74(2.41)    & 21.61(6.70)   & 0.67(0.08)  \\
		RAMP       & 11.38(1.47) & 14.82(2.34)  & 4.61(0.99)   & 8.88(4.35)  & 186.02(51.92)  & 11.85(3.83)   & 0.58(0.07)  \\
		Cox-GEL    & 7.88(4.07)  & 27.46(13.91) & 25.00(21.74) & 6.30(3.26)  & 375.30(189.01) & 68.70(61.24)  & 0.56(0.10)  \\
		Cox-Lasso  & 8.60(1.55)  & 0.18(0.39)   & 4.12(0.04)   & 9.44(1.88)  & 27.46(3.63)    & 4.93(0.04)    & 0.78(0.05)  \\ \bottomrule
	\end{tabular}\label{t:sim2}
\end{table}

\begin{table}[H]
	\caption{Simulation results under the scenarios with AR(0.6): higher-level identification. In each cell, mean (SD) based on 100 replicates.}
	\renewcommand\tabcolsep{2.3pt}
	\begin{tabular}{@{}llllllllll@{}}
		\toprule
		\textbf{Approach} & \textbf{H-M:TP} & \textbf{H-M:FP} & \textbf{H-I:TP} & \textbf{H-I:FP} & & \textbf{H-M:TP} & \textbf{H-M:FP} & \textbf{H-I:TP} & \textbf{H-I:FP} \\ \midrule
		&\multicolumn{4}{c}{$K=100$}  &                                                            & \multicolumn{4}{c}{$K=50$}                                           \\
		\cline{2-5} \cline{7-10}
		& \multicolumn{8}{c}{S1 with censoring rate 40$\%$}                                                                                                           \\
		proposed          & 3.92(0.27)      & 0.00(0.00)      & 1.98(0.14)      & 0.00(0.00)  &    & 3.78(0.58)      & 0.00(0.00)      & 1.34(0.66)      & 0.00(0.00)      \\
		Cox-GEL           & 2.00(1.09)      & 0.00(0.00)      & 0.12(0.33)      & 0.06(0.24)   &   & 1.26(0.72)      & 0.00(0.00)      & 0.12(0.33)   & 0.04(0.28)      \\ \midrule
		&\multicolumn{8}{c}{S2 with censoring rate 40$\%$}                                                                                                           \\
		proposed          & 3.82(0.63)      & 0.00(0.00)      & 1.92(0.34)      & 0.02(0.14)   &   & 3.74(0.63)      & 0.00(0.00)      & 1.28(0.61)      & 0.00(0.00)      \\
		Cox-GEL           & 2.08(1.12)      & 0.00(0.00)      & 0.20(0.45)      & 0.04(0.20)  &    & 1.54(0.84)      & 0.00(0.00)      & 0.08(0.27)      & 0.04(0.28)      \\ \midrule
		&\multicolumn{8}{c}{S3 with censoring rate 40$\%$}                                                                                                           \\
		proposed          & 3.98(0.14)      & 0.00(0.00)      & 1.70(0.46)      & 0.00(0.00)  &    & 3.78(0.55)      & 0.00(0.00)      & 0.36(0.53)      & 0.00(0.00)      \\
		Cox-GEL           & 1.50(0.68)      & 0.00(0.00)      & 0.00(0.00)      & 0.00(0.00)   &   & 1.26(0.56)      & 0.00(0.00)      & 0.00(0.00)      & 0.00(0.00)      \\ \midrule
		&\multicolumn{8}{c}{S1 with censoring rate 20$\%$}                                                                                                           \\
		proposed          & 4.00(0.00)      & 0.00(0.00)      & 2.00(0.00)      & 0.00(0.00)  &    & 4.00(0.00)      & 0.00(0.00)      & 1.80(0.40)      & 0.00(0.00)      \\
		Cox-GEL           & 3.12(1.06)      & 0.00(0.00)      & 0.42(0.61)      & 0.02(0.14)   &   & 2.40(1.20)      & 0.00(0.00)      & 0.08(0.27)      & 0.00(0.00)      \\ \midrule
		&\multicolumn{8}{c}{S2 with censoring rate 20$\%$}                                                                                                           \\
		proposed          & 4.00(0.00)      & 0.00(0.00)      & 2.00(0.00)      & 0.00(0.00)   &   & 3.98(0.14)      & 0.00(0.00)      & 1.86(0.35)      & 0.00(0.00)      \\
		Cox-GEL           & 3.20(1.09)      & 0.00(0.00)      & 0.44(0.61)      & 0.00(0.00)   &   & 2.16(1.02)      & 0.00(0.00)      & 0.10(0.30)      & 0.00(0.00)      \\ \midrule
		&\multicolumn{8}{c}{S3 with censoring rate 20$\%$}                                                                                                           \\
		proposed          & 4.00(0.00)      & 0.00(0.00)      & 1.92(0.27)      & 0.00(0.00)  &    & 4.00(0.00)      & 0.00(0.00)      & 1.00(0.49)      & 0.00(0.00)      \\
		Cox-GEL           & 2.36(1.08)      & 0.00(0.00)      & 0.10(0.30)      & 0.04(0.20)  &    & 1.60(0.83)      & 0.00(0.00)      & 0.00(0.00)      & 0.00(0.00)      \\ \bottomrule
	\end{tabular}\label{sim:t3}
\end{table}

For completeness, we further examine performance of the proposed approach under uncensored settings. Specifically, another six simulation scenarios with completely observed continuous response are designed under the AR(0.6) correlation setting. The summarized results are provided in Tables S9 and S10 in Appendix B of the Supplementary Material. Without censoring, performance of all approaches improves, especially for glinternet and RAMP, but the proposed approach still performs favorably.


\section{Real Data}
We analyze The Cancer Genome Atlas (TCGA) data on cutaneous melanoma (SKCM) and lung adenocarcinoma (LUAD). Data are downloaded from the TCGA Provisional using R package \textit{cgdsr} and the KEGG pathway database. We limit our analysis to the genes that can be mapped to the KEGG pathways.

\subsection{SKCM data}
$18,936$ gene expression measurements are available for 462 patients. Among them, approximately 51.9$\%$ of subjects are censored. The observed survival times are from 0.20 to 369.65 months with the median being 16.99. Since it is expected that the number of genes associated with the survival outcome is not large, we conduct a marginal screening. Specifically, the top 2,000 genes with the smallest p-values computed from a marginal Cox model are selected for downstream analysis. By merging the gene expression data
with the KEGG pathways, 176 pathways with sizes from 1 to 45 are obtained,  containing 1,579 genes in total and 600 distinct genes after removing duplicates.

\begin{figure}[H]
	\centering
	\includegraphics[width=4.5in]{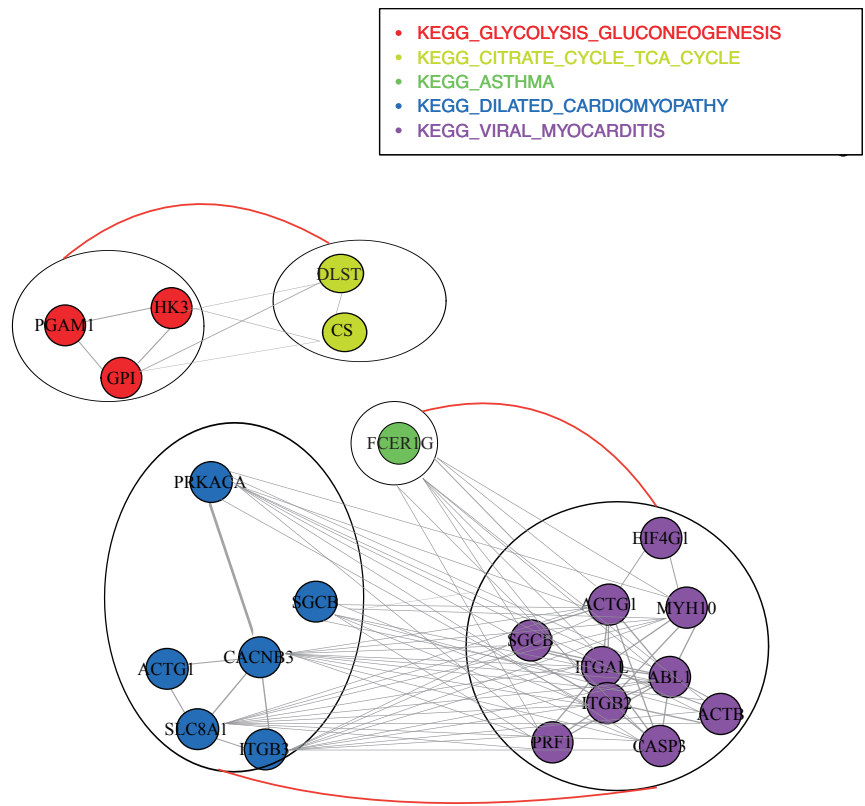}
	\caption{Analysis of the TCGA SKCM data using the proposed approach: identified lower-level genes and gene-gene interactions, as well as higher-level pathways and pathway-pathway interactions. Different colors represent different pathways, and two genes (pathways) are connected if the corresponding interaction is also selected.}
	\label{fig:skcm1}
\end{figure}

The proposed approach identifies 20 distinct genes and 92 distinct interactions (22 genes and 104 interactions before removing duplicates), as well as five KEGG pathways and three pathway-pathway interactions. Figure \ref{fig:skcm1} shows the identified lower-level genes and their interactions, as well as higher-level pathways and pathway-pathway interactions, where an edge represents the identified interaction. The detailed estimation results are provided in the Supplementary Material.

To get a deeper insight into the identified genes, we search the literature for independent evidences of their biological relationships with SKCM. For example, gene CASP3 has been reported to be occasionally mutated in a
wide range of human tumors and cell lines, and this mutation may be accompanied by other alterations in the apoptosis pathway. The abnormal expression and aggregation of ACTB and resulting changes in the cytoskeleton have been shown to be closely associated with a variety of cancers. A high expression level of gene ACTG1 has been found in skin cancer tissues, which contributes to the regulation of cell proliferation and migration through the ROCK signaling pathway. In addition, literature has shown that EIF4G1 is overexpressed in a variety of cancers and plays a vital role in cancer cell survival, thus can be used as a potential target for the treatment of a variety of cancers.

We also examine biological implications of the identified higher-level pathways and pathway-pathway interactions. For example, Citrate cycle (TCA cycle) has been suggested to be significantly down-regulated in tumor formation and progression. Glycolysis has been confirmed to play a significant role in developing metabolic symbiosis in metastatic melanoma progression. Asthma may be associated with a lower risk of cutaneous melanoma. Cardiac metastasis occasionally complicates the progression of tumors, and malignant melanoma has been considered to be with the highest rate of cardiac metastasis.
Myocarditis has been suggested to be an immune-related adverse event with ipilimumab/nivolumab combination therapy for metastatic melanoma. Among the identified pathway-pathway interactions, TCA cycle has been suggested to be linked to glycolysis/gluconeogenesis through phosphoenolpyruvate carboxykinase.
Stress-induced cardiomyopathy has been shown to be a potentially important complication in the course of status asthmaticus, suggesting their interactions. In addition, viral myocarditis has been found to serve as a paradigm for understanding the pathogenesis and treatment of dilated cardiomyopathy. References supporting the above discussions are provided in Appendix B of the Supplementary Material.

We further perform data analysis using alternative approaches. Table \ref{t:realdata} shows the summary comparison results, where the numbers of lower-level main effects and interactions identified by different approaches, their overlaps, as well as RV coefficients are provided. Here, RV coefficient measures the similarity of two data matrices consisting of the identified main effects (interactions), with a larger value indicating a higher similarity. From Table \ref{t:realdata}, significantly different identification results are observed, which is not surprising as these approaches have diverse strategies. With the potentially moderate correlations between genes, variables identified by different approaches have moderate RV coefficients, indicating a certain degree of similarity.

To provide an indirect support for identification results, we examine prediction accuracy based on a resampling technique, where the data is randomly split into a training and a testing set. With 100 resamplings, the mean C-statistics for the testing subjects are 0.537 (proposed), 0.508 (surBayes), 0.504 (glinternet), 0.524 (RAMP), 0.512 (Cox-GEL), and 0.505 (Cox-Lasso). We further consider selection stability for the main effects and interactions identified by each approach and calculate the observed occurrence index (OOI), which is the selected frequency in 100 resamplings. The mean OOI values are 0.821 (proposed), 0.524 (surBayes), 0.351 (glinternet), 0.105 (RAMP), 0.227 (Cox-GEL), and 0.001 (Cox-Lasso). Competitive prediction accuracy and significantly improved selection stability of the proposed approach are observed.

\subsection{LUAD data}
The data contains 18,325 gene expression measurements for 504 subjects, where 322 subjects are censored and the observed survival times lie in [0.13, 238.11] months with median being 15.69. Using a similar prescreening as before, we obtain 180 KEGG pathways consisting of 1,597 genes, among which 605 are distinct.

The proposed approach identifies 32 genes and 179 interactions (50 genes and 337 interactions before removing duplicates), as well as 13 pathways and two pathway-pathway interactions. The graphical representation is presented in Figure \ref{fig:luad} and more detailed estimation results are provided in the Supplementary Material. Independent evidences in the literature indicate our method can recover biologically interesting biomarkers 
that are related to lung cancer. For example, over expression of CTLA4 by lymphocyte subsets might be associated with lung cancer, and thus immunotherapy regimen targeting CTLA4 and Treg cells might be beneficial for lung cancer patients. Gene ITGB1 has been identified as an independent prognostic integrin marker associated with overall survival in NSCLC. In addition, studies have indicated that EIF4G1 acts as an oncoprotein during the development of NSCLC, which may represent a new promising therapeutic target for lung cancer. Gene TPM3, one of the ROS1 fusion partner genes, has been reported in NSCLC. A higher expression level of ITGB4 and lower expression level of CAV1 have been observed in LUAD, which indicates that they may be used as potential prognostic biomarkers. It has also been found that BTK might be an indicator for the status of the tumor microenvironment (TME) in LUAD patients, which provides additional insights for the treatment of LUAD. Moreover, LAMA2 has been reported to be downregulated in LUAD and inhibit LUAD metastasis. CD28 has been demonstrated to be related to poor disease-free survival but long overall survival, and may be involved in the adjustment of the tumor immune microenvironment. Research has shown that PTPRC may be a potential prognostic marker for LUAD, and it may affect the function of immune cells by participating in the regulation of TME immune status.

\begin{figure}[H]
	\centering
	\includegraphics[width=4.5in]{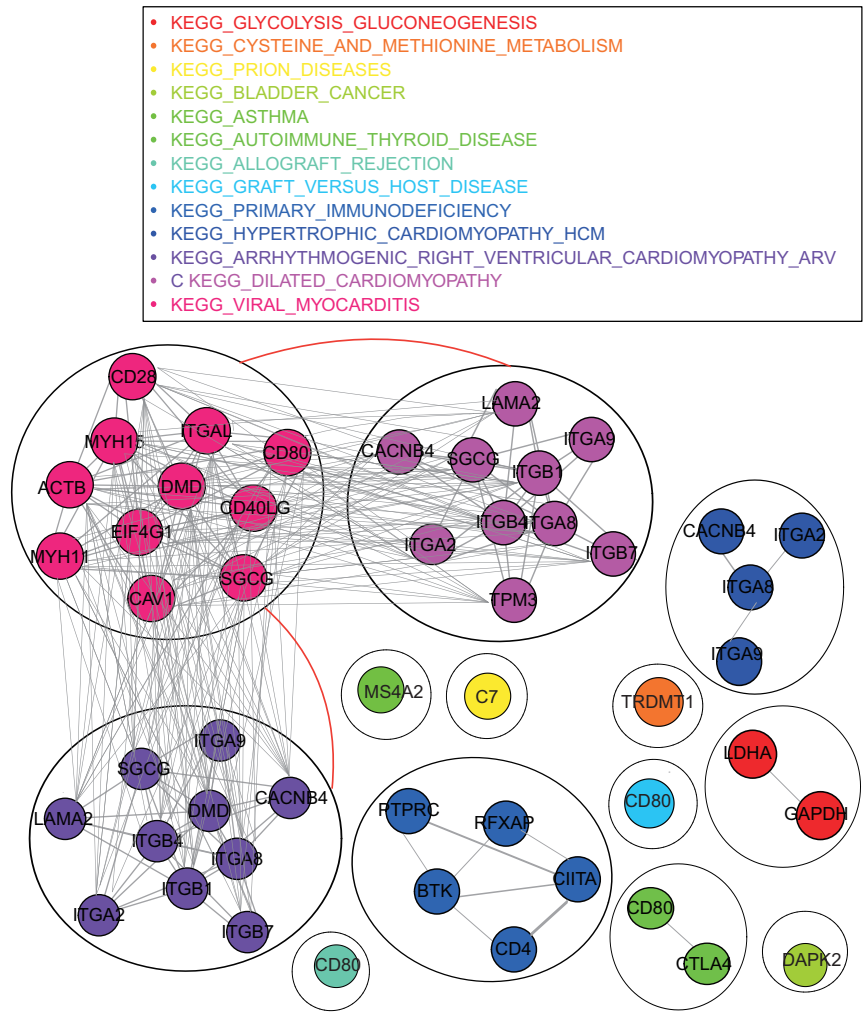}
	\caption{Analysis of the TCGA LUAD data using the proposed approach: identified lower-level genes and gene-gene interactions, as well as higher-level pathways and pathway-pathway interactions. Different colors represent different pathways, and two genes (pathways) are connected if the corresponding interaction is also selected.}
	\label{fig:luad}
\end{figure}

In addition, the identified higher-level pathways and pathway-pathway interactions have possible associations with lung cancer. For instance, studies have shown that as a reverse glycolysis pathway, gluconeogenesis can generate glucose from small carbohydrate precursors, which is crucial for the growth of tumor cells. Researchers have found that the metabolism of cysteine and methionine involves the role of target genes in the molecular mechanism of NSCLC. Primary immunodeficiency pathway has been correlated with the early relapse of patients with early-stage LUAD. In addition, cardiac comorbidity has been suggested as an independent risk factor for radiation pulmonary toxicity in patients with lung cancer. For pathway-pathway interactions, we notice that pathways dilated cardiomyopathy, viral myocarditis and ARVC are all related to heart disease. Specifically, studies have shown that there is a significant increase in circulating cytokines in patients with myocarditis and cardiomyopathy. Moreover, up to 20\% of myocarditis patients may subsequently develop chronic inflammatory dilated cardiomyopathy. These relationships may suggest the potential importance of their interactions for lung cancer. We refer to Appendix B in the Supplementary Material for the references.

\begin{table}[H]
	\caption{Data analysis: numbers of lower-level main effects and interactions (diagonal elements) identified by different approaches and their
		overlaps and RV coefficients (off-diagonal elements).}
	\begin{tabular}{@{}clllllll@{}}
		\toprule
		\textbf{SKCM}                         & Approach   & proposed  & surBayes  & glinternet & RAMP      & Cox-GEL   & Cox-Lasso \\ \midrule
		\multirow{6}{*}{\textbf{Main}}        & proposed   & 20        &
		0(0.36)   & 5(0.30)    & 0(0.23)   & 0(0.32)   & 2(0.47)   \\
		& surBayes   &    & 2         &1(0.16)&	0(0.11)&	1(0.16)	&0(0.19)
		\\                  & glinternet &    &    & 215        & 37(0.79)	&11(0.27)&	2(0.21)  \\
		& RAMP       &    &    &     & 53        & 1(0.13)	&1(0.17)
		\\
		& Cox-GEL    &    &    &     &    & 28        & 0(0.24)   \\
		& Cox-Lasso  &    &    &     &    &    & 5         \\ \midrule
		\multirow{6}{*}{\textbf{Interaction}} & proposed   & 92        & 0(0.32)   & 2(0.04)    & 0(0.10)   & 0(0.40)   & 0(0.20)   \\
		& surBayes   &    & 151       & 0(0.34)&	0(0.18)	&0(0.10)&	0(0.43)   \\
		& glinternet &    &    & 174        & 0(0.39)&	1(0.01)	&3(0.71)
		\\
		& RAMP       &    &    &     & 160       &0(0.03)	&0(0.36)
		\\
		& Cox-GEL    &    &    &     &    & 326       & 0(0.06)   \\
		& Cox-Lasso  &    &    &     &    &    & 104       \\ \midrule
		\textbf{LUAD}                         &            &           &           &            &           &           &           \\ \midrule
		\multirow{6}{*}{\textbf{Main}}        & proposed   & 32        & 0(0.30)&	13(0.58)&	2(0.38)&	2(0.45)&	0(0.17)
		\\
		& surBayes   &    & 5         & 4(0.32)&	1(0.21)	&0(0.18)&	1(0.46)
		\\
		& glinternet &    &    & 182 &     33(0.75)&	8(0.50)&	4(0.27)
		\\
		& RAMP       &    &    &     & 49        &0(0.35)&	2(0.19)
		\\
		& Cox-GEL    &    &    &     &    & 18        & 0(0.23)  \\
		& Cox-Lasso  &    &    &     &    &    & 7         \\ \midrule
		\multirow{6}{*}{\textbf{Interaction}} & proposed   & 179       & 0(0.23)&	0(0.14)&	0(0.02)&	1(0.27)	&0(0.13)
		\\
		& surBayes   &    & 199       & 0(0.12)&	0(0.02)	&1(0.38)&	0(0.12)	
		\\
		& glinternet &    &    & 150        & 1(0.07)&	0(0.14)&	0(0.11)
		\\
		& RAMP       &    &    &     & 124       &0(0.02)&	0(0.03)
		\\
		& Cox-GEL    &    &    &     &    & 153       & 1(0.14)   \\
		& Cox-Lasso  &    &    &     &    &    & 96        \\ \bottomrule
	\end{tabular}\label{t:realdata}
\end{table}

We also compare the proposed approach with alternatives. The summarized results are presented in Table \ref{t:realdata}, from which we conclude similarly that the identification results vary significantly across approaches but have moderate RV coefficients. Based on 100 resamplings, we compute the mean C-statistic and OOI values as (0.534, 0.987) for the proposed approach, (0.502, 0.563) for surBayes, (0.512, 0.284) for glinternet, (0.524, 0.098) for RAMP, (0.527, 0.207) for Cox-GEL, and (0.504, 0.001) for Cox-Lasso,  respectively. Again, the proposed approach has favorable prediction performance and selection stability.

\section{Discussion}
Genetic interaction analysis for modelling the outcomes of complex diseases, especially for prognostic survival, is still a nontrivial task. The accumulated pathway information provides us with more opportunities to search for important interactions more effectively. Extensive literature has shown that pathways are related, and their interactions are also conducive to the prognosis of complex diseases. We have developed a new interaction analysis approach, where the two-level selection has been considered. The advancement of the proposed approach lies in its creative use of Bayesian technique and ingeniously accommodating censoring data. In addition, advancing from existing single-level interaction analysis, effective priors have been proposed to explore two-level selection and M-I hierarchy. The proposed approach has been formulated as a hybrid Bayesian model with a tractable posterior, making the variational Bayesian EM algorithm possible. The analysis of simulated data has demonstrated that the proposed approach outperforms the alternatives on identification, estimation, and prediction accuracy by a large margin.

We have also applied it to two TCGA studies. It has been observed that the survival time of these two types of cancer is determined by the two-level factors, many of which have been proved to play an important role in the development of cancer. Interestingly, the identified pathways are either metabolism-related or disease-related, among which only those in the same type of pathways have important interactions. The identified pathway-pathway interactions may provide new evidence for the importance of the affiliated gene-gene interactions that have not yet been reported in the literature. In addition, the proposed approach has been observed to have better prediction accuracy and selection stability. In our data analysis, we have focused on gene expression data. It will be of interest to include additional confounders, such as clinical and demographic factors. As they usually have a low dimension and are pre-selected, as well as without a two-level structure, their coefficients will not be subject to two-level selection. It is also worthwhile to mention that the proposed approach can be applied to other data with a two-level structure and censoring, enjoying very broad applicability.

\section*{Acknowledgements}
The second author was supported in part by National Institutes of Health [CA204120]. The third author was supported in part by
National Natural Science Foundation of China [12071273]; ``Chenguang Program'' supported by Shanghai Education Development Foundation and Shanghai Municipal Education Commission [18CG42]; and Shanghai Pujiang Program [19PJ1403600].

\bibliography{refajust}

\section*{SUPPLEMENTARY MATERIAL}

	Supplementary Material to ``Two-level Bayesian interaction analysis for survival data incorporating pathway information''. We provide the detailed derivation of the proposed two-level Bayesian interaction analysis model, details of the VBEM algorithm, weighted method for the use of glinternet and RAMP approaches in survival data, and additional results of simulations and data analysis. In addition, we provide the coefficient estimation results for real data using the proposed approach (RealData.xlsx).

\end{document}